\documentclass[aps,prd,12pt,notitlepage,preprintnumbers,showpacs,nofootinbib,
tightenlines]{revtex4-1}
\usepackage[pdftex]{graphicx}
\usepackage{amsmath}
\usepackage{amssymb}
\usepackage{hyperref}
\usepackage{bm}
\usepackage{slashed}
\usepackage{times}
\usepackage{color}
\usepackage{ulem}
\definecolor{Red}{rgb}{1.,0.,0.}

\definecolor{Blue}{rgb}{0.,0.,1.}

\definecolor{nicered}{rgb}{0.7,0.1,0.1}
\definecolor{nicegreen}{rgb}{0.1,0.5,0.1}
\hypersetup{colorlinks,citecolor=nicegreen,linkcolor=nicered}

\begin{document}

\newcommand{\beq}{\begin{eqnarray}}
\newcommand{\eeq}{\end{eqnarray}}
\newcommand{\ben}{\begin{enumerate}}
\newcommand{\een}{\end{enumerate}}
\newcommand{\be}{\begin{equation}}
\newcommand{\ee}{\end{equation}}

\newcommand{\non}{\nonumber\\ }

\newcommand{\jpsi}{J/\Psi}

\newcommand{\ppa}{\phi_\pi^{\rm A}}
\newcommand{\ppp}{\phi_\pi^{\rm P}}
\newcommand{\ppt}{\phi_\pi^{\rm T}}
\newcommand{\ov}{ \overline }

\newcommand{\zerot}{ {\textbf 0_{\rm T}} }
\newcommand{\kt}{k_{\rm T} }
\newcommand{\fb}{f_{\rm B} }
\newcommand{\fk}{f_{\rm K} }
\newcommand{\rk}{r_{\rm K} }
\newcommand{\mb}{m_{\rm B} }
\newcommand{\mw}{m_{\rm W} }
\newcommand{\im}{{\rm Im} }

\newcommand{\kks}{K^{(*)}}
\newcommand{\acp}{{\cal A}_{\rm CP}}
\newcommand{\pb}{\phi_{\rm B}}

\newcommand{\xeba}{\bar{x}_2}
\newcommand{\xsba}{\bar{x}_3}
\newcommand{\peas}{\phi^A}

\newcommand{\pvsl}{ p \hspace{-2.0truemm}/_{K^*} }
\newcommand{\esl}{ \epsilon \hspace{-2.1truemm}/ }
\newcommand{\psl}{ p \hspace{-2truemm}/ }
\newcommand{\ksl}{ k \hspace{-2.2truemm}/ }
\newcommand{\lsl}{ l \hspace{-2.2truemm}/ }
\newcommand{\nsl}{ n \hspace{-2.2truemm}/ }
\newcommand{\vsl}{ v \hspace{-2.2truemm}/ }
\newcommand{\epsl}{\epsilon \hspace{-1.8truemm}/\,  }
\newcommand{\bfkk}{{\bf k} }
\newcommand{\calm}{ {\cal M} }
\newcommand{\calh}{ {\cal H} }
\newcommand{\calo}{ {\cal O} }

\def \appb{{\bf Acta. Phys. Polon. B }  }
\def \cpc{ {\bf Chin. Phys. C } }
\def \ctp{ {\bf Commun. Theor. Phys. } }
\def \epjc{{\bf Eur. Phys. J. C} }
\def \ijmpcs{{\bf Int. J. Mod. Phys. Conf. Ser.} }
\def \jhep{{\bf J. High Energy Phys. } }
\def \jpg{ {\bf J. Phys. G} }
\def \mpla{{\bf Mod. Phys. Lett. A } }
\def \npb{ {\bf Nucl. Phys. B} }
\def \plb{ {\bf Phys. Lett. B} }
\def \ppn{ {\bf Phys. Part. Nucl. } }
\def \ppnp{{\bf Prog.Part. Nucl. Phys.  } }
\def \pr{  {\bf Phys. Rep.} }
\def \prc{ {\bf Phys. Rev. C }}
\def \prd{ {\bf Phys. Rev. D} }
\def \prl{ {\bf Phys. Rev. Lett.}  }
\def \ptp{ {\bf Prog. Theor. Phys. }}
\def \zpc{ {\bf Z. Phys. C}  }
\def \jpg{ {\bf J.Phys.-G-}  }
\def \ap{ {\bf Ann. of Phys}  }

\title{Rho-pion transition form factors in the $k_T$ factorization formulism revisited}
\author{Jun Hua$^{1}$}
\author{Shan Cheng$^{2}$} \email{chengshan-anhui@163.com}
\author{Ya-lan Zhang$^{3}$}
\author{Zhen-Jun Xiao$^{1,4}$ } \email{xiaozhenjun@njnu.edu.cn}
\affiliation{1.  Department of Physics and Institute of Theoretical Physics,
Nanjing Normal University, Nanjing, Jiangsu 210023, People's Republic of China,}
\affiliation{2. School of Physics and Microelectronics Science, Hunan University, 
Changsha, Hunan 410082, People's Republic of China,}
\affiliation{3. Department of Faculty of Mathematics and Physics, Huaiyin Institute of Technology, 
Huaian, Jiangsu 223001, People's Republic of China,}
\affiliation{4. Jiangsu Key Laboratory for Numerical Simulation of Large Scale Complex Systems,
Nanjing Normal University, Nanjing, Jiangsu 210023, People's Republic of China}
\date{\today}
\begin{abstract}

We revisit the evaluations for the spacelike and timelike $\rho \pi$ transition form factors 
$F_{\rho\pi}(Q^2)$ and $G_{\rho\pi}(Q^2)$ with the inclusion of the the next-to-leading order (NLO) 
QCD contributions in the framework of the $k_T$ factorization theorem.
The infrared divergence is regularized by the transversal momentum carried by external valence 
quarks, and ultimately absorbed into the meson wave functions.
In the region of  $ Q^2  \leqslant 2  \, \textrm{GeV}^2 $, where PQCD factorization apporach 
applicable, the NLO contribution can bring no larger than $ 35\%$ enhancement to the 
spacelike form factor $F_{\rho\pi}(Q^2)$.
For the timelike form factor derived under the kinematic exchanging symmetry,
this contribution is also under control when the momentum transfer squared is large enough.
We also prolong our prediction into the small $Q^2$ region by taking the Lattice QCD results into 
account, and subsequently obtain the coupling $g_{\rho\pi\gamma} = G_{\rho\pi}(0)=0.596$.
\end{abstract}



\maketitle

\section{Introduction}
Rho-pion transition form factor carries the information of momentum redistribution between all the constituents in initial and final states,
when a photon is hitting on one constituent and the bound state does not fall apart\cite{IoffeQB}.
This physical quantity, in principle, is evoluted in the whole momentum transfer squared extent,
but actually, from the traditional QCD based approaches, we can only calculate it in the intermediate and
large energy regions due to the color confinement \cite{KhodjamirianTK,ZuoHZ,YuHP,ZhangMXA},
while in the small energy scope, it can been investigated only in lattice QCD
\cite{FengGBA,OwenGVA,OwenFRA,BricenoKKP} and measured in experiments \cite{HustonWI}.

The perturbative QCD (PQCD) approach was initially proposed to calculate pion
electromagnetic form factor\cite{BottsKF}
with the well done resummation technique eliminating endpoint divergence \cite{LiNU},
And recently, this work has stepped forward to next-to-leading-order (NLO) QCD corrections
\cite{LiNN,ChengGBA}.
the result turns out that the convergency of perturbative expansion is very good in the corresponding energy region.
The basic idea is to keep the transversal momentum of external valence quarks in the denominates of internal propagates,
and drop the transversal momentum emerged in the numerator,
because these terms bring the gauge dependence which should be compensated with the soft gluon correction
(three-parton distribution amplitude contribution) \cite{ChenPN}
due to the gauge invariant matrix element at sub-leading power correction\cite{QiuDN},
the infrared regulators obtained in this way are single logs of transversal momentum squared and is
absorbed completely into the definition of nonperturbative meson wave function.

Factorization of the similar exclusive process, says rho-pion transition, is also derived both in light-cone collinear approach\cite{AnikinBF}
and in PQCD approach\cite{ChengRRA} up to sub leading twist.
Following the leading-order calculation of spacelike form factor\cite{ZhangMXA}, we focus on the NLO correction in this paper,
and use the kinematic exchanging symmetry between positive and negative energy axises to study the timelike form factor.
By taking into account the lattice result in the small energy region, where PQCD is invalid,
we interpolate the form factor in the whole energy interval and try to
determine the rho-pion coupling $g_{\rho\pi\gamma}$.
We note here that only the vector current $J_{\mu,\vert \lambda \vert =1}$ accompanied by the transversal polarized rho meson gives nonzero contribution to rho-pion transition,
the residual $\gamma_5$ existed in hadron matrix element make it is very different from the pion form factor,
says the spacelike and timelike rho-pion matrix element have the same expression in terms of corresponding form factors.

This paper is organized as follows. In the following section we briefly summarize the LO prediction of rho-pi form factor from PQCD approach.
In Sec.~III, the NLO correction to form factor is calculated, along with the discussion of infrared divergence.
Numerics is performed in Sec.~IV, we parameterize rho-pion form factor in the full spacelike energy region with the lattice result at small energy points.
Sec.~V contains the conclusion.

\section{Rho-pion form factor at Leading Order}

\begin{figure}[htbp]
\begin{center}
\vspace{-2cm}
\includegraphics[width=0.8\textwidth]{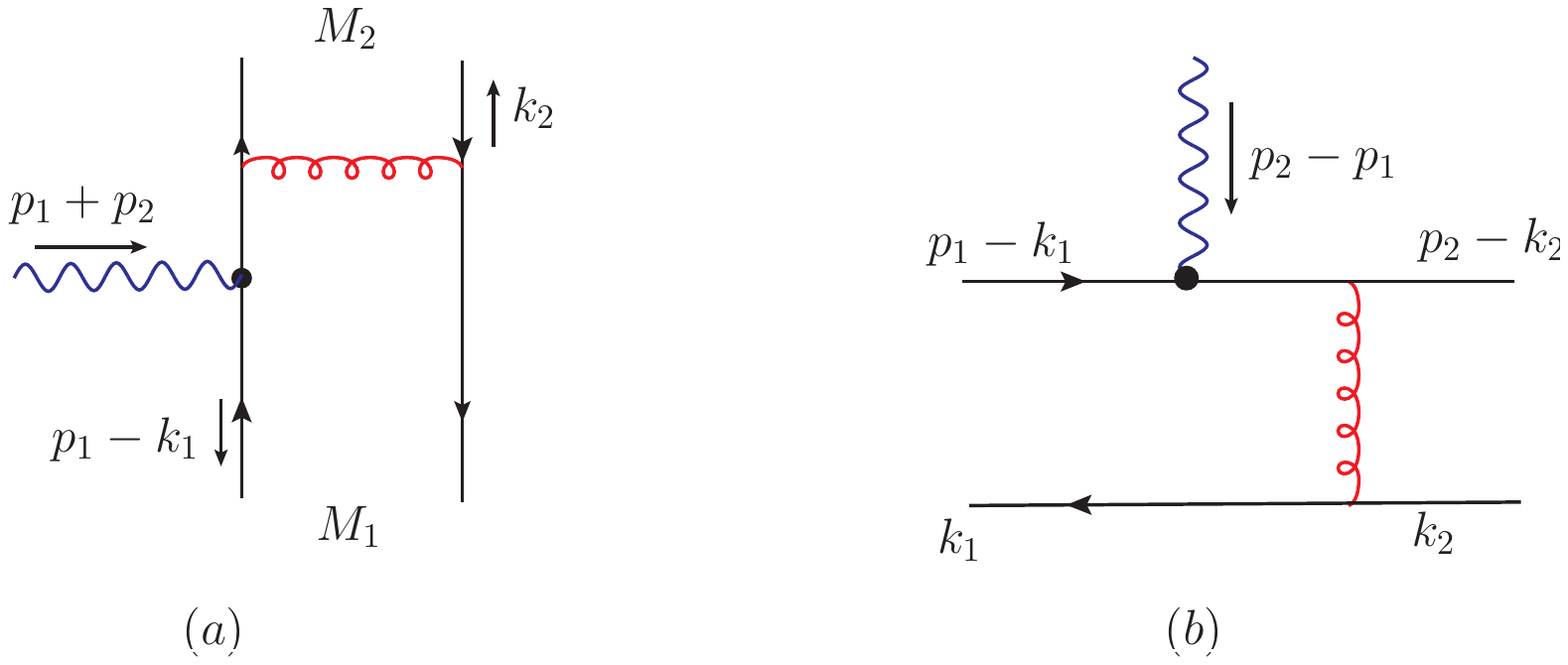}
\vspace{-12.8cm}
\caption{Feynman diagrams for timelike (a), and spacelike (b) rho-pion transition at LO.}
\label{fig:fig1}
\end{center}
\end{figure}

Distinguishing by the momentum transfer carried by vector current, timelike and spacelike rho-pion transitions at LO are plotted in Fig.~\ref{fig:fig1}.
There are three other diagrams for each type form factor, with the virtual photon current locating on the other three quark/antiquark lines.
We definitely take $M_1$ as $\rho^+$ and $\rho^-$ for Fig.~\ref{fig:fig1}(a) and (b), respectively, and $M_2$ is $\pi^-$ for both diagrams,
$M_1$ carries the "positive" momentum $p_1=\frac{Q}{\sqrt{2}}(1,\gamma^2_{\rho},0)$,
while $M_2$ carries the "negative" momentum $p_2=\frac{Q}{\sqrt{2}}(\gamma^2_{\pi},1,0)$ along the light cone,
with the dimensionless $\gamma^2_{\rho,\pi}\equiv M^2_{\rho,\pi}/Q^2 $.
The anti-quark $\bar{d}$ in initial $\rho^-$ and final $\pi^-$ carries momentum $k_1=(x_1p^+_1,0,\mathbf{k_{1T}})$ and $k_2=(0,x_2p^-_2,\mathbf{k_{2T}})$, respectively,
while in the final $\rho^+$ the momentum fraction $x_1$ is carried by quark $u$, $\mathbf{k_T}$ represents the transversal momentum.
In this convention, momentum transfer squared in timelike and spacelike transition is $Q^2=(p_1+p_2)^2$ and $q^2=(p_1-p_2)^2$, respectively,
$Q^2 = -q^2$ in the large momentum limit.
The related meson wave functions are written as,
{\small
\beq
&&\langle 0 \vert \bar{u}(0)_j d(z_1)_l \vert \rho^-(p_1,\epsilon_{T}) \rangle = \frac{1}{\sqrt{2N_C}} \int_0^1 dx_1 e^{ix_1p_1z_1}
\left\{\psl_1 \esl_T \phi_\rho^T(x_1) + m_\rho \esl_T \phi^v_\rho(x_1) \right.  \non
&&\left. \hspace{8cm} + m_\rho i \epsilon_{\mu\nu\rho\sigma} \gamma^\mu \gamma_5 \epsilon^\nu_T n^\rho v^\sigma \phi_\rho^a(x_1)  \right\}_{lj} , \,\,\,
\label{eq:1}\\
&&\langle \pi^-(p_2) \vert \bar{d}(z_2)_j u(0)_l \vert 0 \rangle = \frac{i}{\sqrt{2N_C}} \int_0^1 dx_2 e^{ix_2p_2z_2}  \gamma_5
\left\{\psl_2 \phi_\pi(x_2) + m_0^\pi \phi^P_\pi(x_2) \right.  \non
&&\left. \hspace{8cm} + m_0^\pi (\vsl \nsl-1) \phi_\pi^T(x_2)  \right\}_{lj} , \,\,\,
\label{eq:2}\\
&&\langle \rho^+(p_1,\epsilon_{T}) \vert \bar{u}(z_1)_j d(0)_l \vert  0 \rangle = \frac{1}{\sqrt{2N_C}} \int_0^1 dx_1 e^{ix_1p_1z_1}
\left\{\esl_T \psl_1  \phi_\rho^T(x_1) + m_\rho \esl_T \phi^v_\rho(x_1) \right.  \non
&&\left. \hspace{8cm} + m_\rho i \epsilon_{\mu\nu\rho\sigma} \gamma_5 \gamma^\mu \epsilon^\nu_T n^\rho v^\sigma \phi_\rho^a(x_1)  \right\}_{lj} ,
\label{eq:3}
\eeq}\\
where $\phi_\pi$ and $\phi^T_\rho$ denote the twist-2 distribution amplitudes (DAs), $\phi^{P,T}_\pi$ and $\phi^{v,a}_\rho$ are twist-3 DAs,
dimensionless vectors $n=(1,0,\mathbf0_T)$ and $v=(0,1,\mathbf0_T)$, $N_c$ is the number of colors.
Rho-pion transition matrix element is then formulated in terms of form factor associated with the antisymmetry tensor,
{\small
\beq
&&\langle \pi^-(p_2) \vert \, J_{\mu,\vert\lambda\vert=1}(p_1-p_2) \,  \vert \rho^-(p_1,\epsilon_T) \rangle
= ie \mathcal{F}_{\rho\pi}(Q^2) \epsilon_{\mu\nu\rho\sigma} \epsilon^\nu_T n^\rho v^\sigma p_1^+p_2^- , \,\,\,
\label{eq:4}\\
&& \langle \rho^{+}(p_1,\epsilon) \pi^-(p_2) \vert \, J_{\mu,\vert\lambda\vert=1}(p_1+p_2) \, \vert 0 \rangle
= ie \mathcal{G}_{\rho\pi}(Q^2) \epsilon_{\mu\nu\rho\sigma} \epsilon^\nu_T v^\rho n^\sigma p_1^+p_2^- ,
\label{eq:5}
\eeq}\\
where $J_\mu = \frac{2}{3}e \bar{u}\gamma_\mu u -\frac{1}{3}e \bar{d}\gamma_\mu d $ is the electromagnetic current .

We derive the spacelike rho-pion transition form factor up to subleading twist in three terms corresponding to different Dirac structures of initial and final meson states,
{\small
\beq
&&\mathcal{F}^{LO}_{\rho\pi}(Q^2)=\frac{64\pi}{9}  \alpha_s(\mu_f) \int_0^1 dx_1dx_2 \int_0^\infty b_1db_1 b_2db_2 \, \textrm{exp}[-S_{\rho\pi}(x_i,b_i,Q,\mu)] \non
&&\hspace{2cm} \left\{ m_\rho (\phi_\rho^v(x_1)-\phi_\rho^a(x_1)) \phi_\pi^A(x_2) h(x_2,x_1,b_2,b_1) \right. \non
&&\left. \hspace{2cm} + x_1 m_\rho (\phi_\rho^v(x_1)-\phi_\rho^a(x_1)) \phi_\pi^A(x_2) h(x_1,x_2,b_1,b_2) \right. \non
&&\left. \hspace{2cm} + 2m_0^\pi \phi_\rho^T(x_1) \phi^P_\pi(x_2) h(x_1,x_2,b_1,b_2)  \right\} S_t(x_1) S_t(x_2),
\label{eq:6}
\eeq}\\
in which, due to the chiral enhancement and end-point effect,
the third term with twist-2 rho DAs and twist-3 pion DAs gives the dominate contribution, shows $\geqslant 90\%$,
that's why in the following we concentrate only on this term for the NLO gluon radiative correction.
$S_t(x)$ is the threshold resummation function parameterized in the simple power-function formula\cite{StermanAJ,CataniNE,LiGI,WeiFM},
$S_{\rho\pi}$ is the $k_T$ Sudakov factor for the transversal momentum\cite{LiNU,LiIS,KurimotoZJ}.
The hard function $h(x_1,x_2,b_1,b_2)$ is obtained from the Fourier transfer of propagators on transversal components.

Timelike rho-pion form factor $\mathcal{G}^{(LO)}_{\rho\pi}(Q^2)$ can be obtained in the similar way with substitute $x_1 \leftrightarrow -x_1$,
which subsequently lead to the replacement $h(x_1,x_2,b_1,b_2) \rightarrow h'(x_1,x_2,b_1,b_2)$.
{\small
\beq
&&h(x_1,x_2,b_1,b_2)=K_0(\sqrt{x_1x_2}Qb_2) \left[ \Theta(b_1-b_2) I_0(\sqrt{x_1}Qb_2) K_0(\sqrt{x_1}Qb_1) + b_1 \leftrightarrow b_2 \right], \,\,\, \non
&&h'(x_1,x_2,b_1,b_2)= K_0(i\sqrt{x_1x_2}Qb_2) \left[ \Theta(b_1-b_2) I_0(i \sqrt{x_1}Qb_2) K_0(i \sqrt{x_1}Qb_1) + b_1 \leftrightarrow b_2 \right].
\label{eq:8}
\eeq}\\
We emphasis the exchanging symmetry of the internal propagators between the spacelike and timelike one\cite{ZhangMXA},
which is the basic argument we used to derive the NLO timelike rho-pion form factor.

\section{Next-to-leading-order correction to Rho-pion form factor}

In this section we consider the NLO gluon radiative correction to rho-pion transition form factor,
we firstly calculate the correction to spacelike form factor in the framework of $k_T$ dependent factorization,
and using the kinematic exchanging symmetry to derive the NLO timelike form factor.
Considering the Sudakov effect to the $\bar{q}q$ bound states, here are rho and pion mesons, our calculation is based on the follow hierarchy\cite{LiNN,ChengGBA},
{\small
\beq
Q^2 \gg x_1Q^2 \thicksim x_2Q^2 \gg x_1x_2Q^2 \gg k^2_{1T} \thicksim k^2_{2T}.
\label{eq:9}
\eeq}\\

\subsection{Spacelike rho-pion form factor at NLO }
The NLO hard kernel in $k_T$ factorization theorem is defined by taking the difference between full amplitude and
effective amplitude, where the wave functions in the latter one absorb all infrared (IR) divergence at a certain order of strong coupling,
{\small
\beq
H^{(1)}(x_1,k_{1T},x_2,k_{2T},Q^2) &=& G^{(1)}(x_1,k_{1T},x_2,k_{2T},Q^2) \non
&-& \int{dx^{'}_1 d^{2} k^{'}_{1T} \, \mathbf{\Phi}^{(1)}_{I}(x_1,k_{1T};x^{'}_1,k^{'}_{1T})} \, \mathcal{H}^{(0)}(x^{'}_1,k^{'}_{1T},x_2,k_{2T},Q^{2}) \non
&-& \int{dx^{'}_2d^{2}k^{'}_{2T} \, \mathcal{H}^{(0)}(x_1,k_{1T},x^{'}_2,k^{'}_{2T},Q^2) \, \mathbf{\Phi}^{(1)}_{F}(x^{'}_2,k^{'}_{2T};x_2,k_{2T})}.
\label{eq:10}
\eeq}\\
$\Phi^{(1)}_I,\Phi^{(1)}_F$ presents the $O(\alpha_s)$ initial and final wave function with the integrated loop momentum flowing in, respectively.
When the loop moment does not flow in, $\mathcal{H}^{(0)}$ is exactly the LO hard kernel in our interesting
{\small
\beq
H^{(0)}(x_1,k_{1T},x_2,k_{2T},Q^2) = \frac{64 \pi \alpha_s(\mu)}{9} \frac{2 m_0^\pi \phi_\rho^T(x_1) \phi_\pi^P(x_2)}{(k_1-k_2)^2 (p_2-k_1)^2} \, .
\label{eq:11}
\eeq}\\
In case the loop momentum flowing in, the momentum of constitutes in $\mathcal{H}^{(0)}$ should redistribute,
which lead to the modified momentum fraction $\delta(x'_1-x_1+l^+/p_1^+) \, \delta(k'_{1T}-k_{1T}+l_T)$ and $\delta(x'_2-x_2+l^-/p_2^-) \, \delta(k'_{2T}-k_{2T}+l_T)$.

\subsubsection{Full amplitudes at NLO}

The full amplitudes at NLO, according to the degree of complexity, include the self correction, vertex correction,
box and pentagon  correction; or in other words, the calculation corresponds to two-point,
three-point, four-point integral, respectively. We define the dimensionless ratios
{\small
\beq
\delta_1=\frac{k^2_{1T}}{Q^2},\quad
\delta_2=\frac{k^2_{2T}}{Q^2},\quad
\delta_{12}=\frac{-(k_{1}-k_{2})^2}{Q^2} \, .
\label{eq:12}
\eeq}\\
In this way, the soft and collinear divergences are both regulated by $\ln\delta_i$, and their overlap singularity is regulated by double log $\ln^2\delta_i$.
The ultraviolet (UV) poles, which is not the focal point in this paper, are processed in dimensional regulation (regulated by $1/\varepsilon$) and
redefined in the $\overline{\mathrm{MS}}$ scheme.

\begin{figure}[htbp]
\begin{center}
\vspace{0cm}
\includegraphics[width=0.6\textwidth]{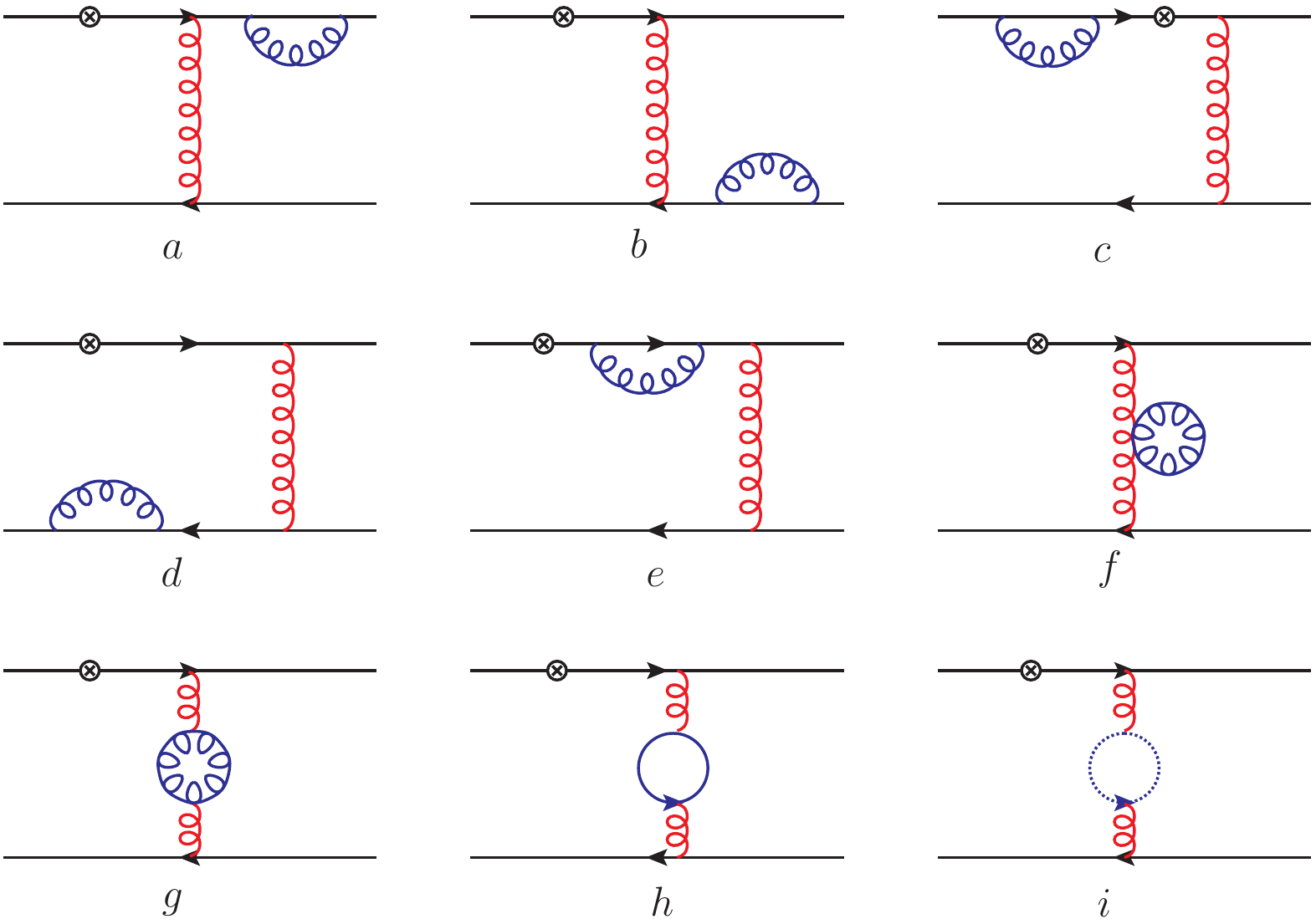}
\vspace{0.5cm}
\caption{Self-energy corrections to Fig.~\ref{fig:fig1}(b).}
\label{fig:fig2}
\end{center}
\end{figure}


The simplest correction include the quark and gluon self-energy correction, as shown in Fig.~\ref{fig:fig2},
whose amplitudes are reducible since the integral momentum does not pollute the hard kernel.
{\small
\beq
G^{(1)}_{2a+2b+2c+2d}&=&-\frac{\alpha_s C_F}{4\pi}
\left [\frac{2}{\varepsilon}+\ln\frac{4\pi \mu^2}{\delta_2Q^2 e^{\gamma_E}}+\ln\frac{4\pi \mu^2}{\delta_1Q^2 e^{\gamma_E}}+4 \right]H^{(0)}  \, ,
\label{eq:13} \\
G^{(1)}_{2e}&=&-\frac{\alpha_s C_F}{4\pi}\left [\frac{1}{\varepsilon}+\ln\frac{4\pi \mu^2}{x_1Q^2 e^{\gamma_E}}+2 \right ]H^{(0)} \, ,
\label{eq:14}\\
G^{(1)}_{2f+2g+2h+2i}&=&\frac{\alpha_s C_F}{4\pi}(\frac{5}{3}N_c-\frac{2}{3}N_f)
\left [\frac{1}{\varepsilon}+\ln \frac{4\pi \mu^2}{\delta_{12}Q^2 e^{\gamma_E}}+2\right ]H^{(0)} \, ,
\label{eq:15}
\eeq}\\
where $\mu $ is renormalization scale, $\gamma_E$ is Euler constant, $N_f$ is the number of quark flavors.

\begin{figure}[htbp]
\begin{center}
\vspace{0cm}
\includegraphics[width=0.6\textwidth]{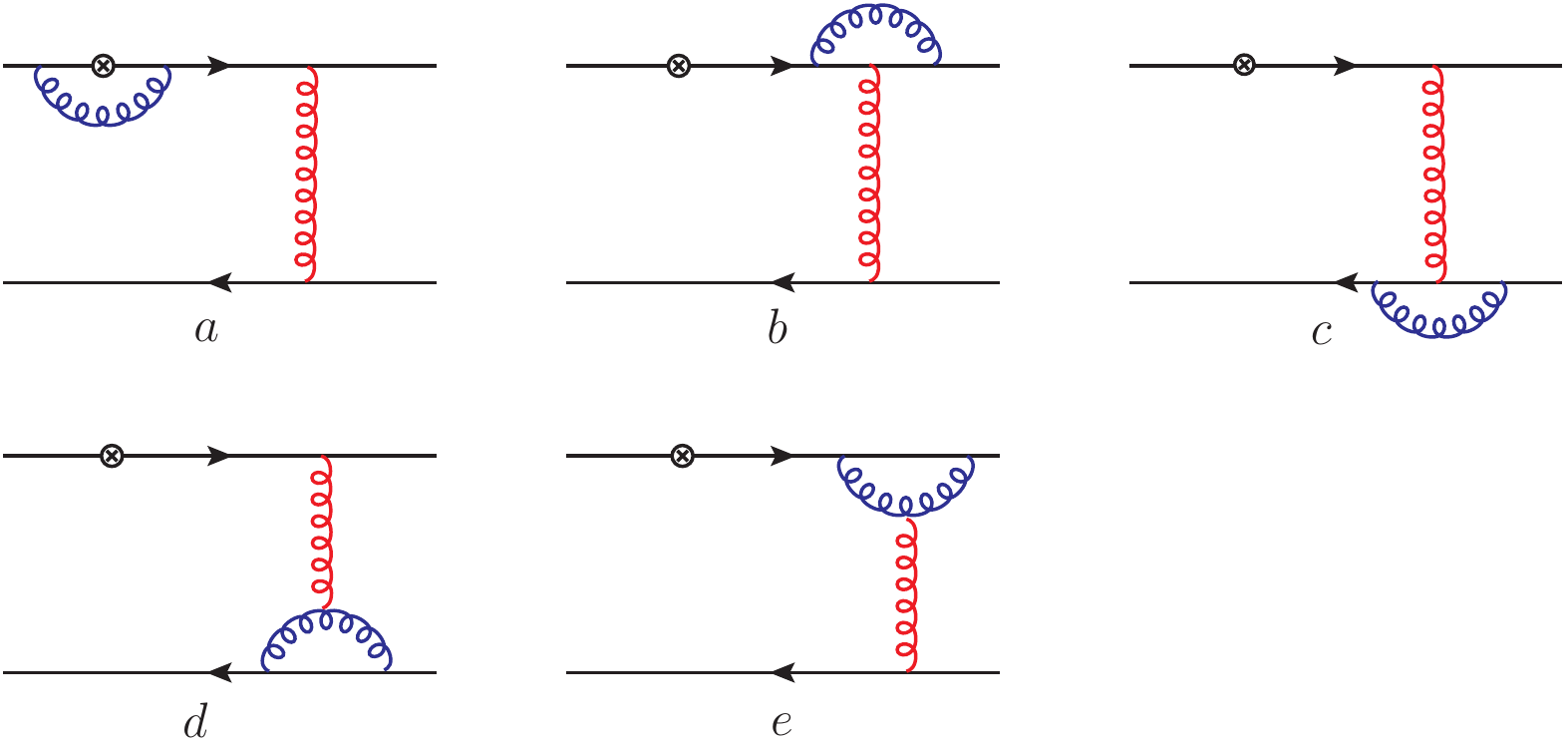}
\vspace{0.5cm}
\caption{Vertex corrections to Fig.~\ref{fig:fig1}(b).}
\label{fig:fig3}
\end{center}
\end{figure}


Calculating the vertex diagrams depicted in Fig.~\ref{fig:fig3} results in the following results:
{\small
\beq
G^{(1)}_{3a}&=&\frac{\alpha_s C_F}{4\pi}\left [\frac{1}{\varepsilon}+\ln\frac{4\pi \mu^2}{Q^2 e^{\gamma_E}}-
2\ln{\delta_1}\ln{x_1}-2\ln{\delta_1}-2\ln{x_1}-\frac{\pi^2}{3}+\frac{3}{2} \right ] H^{(0)} \, ,
\label{eq:16}\\
G^{(1)}_{3b}&=&-\frac{\alpha_s}{8\pi N_c}
\left [\frac{1}{\varepsilon}+\ln\frac{4\pi \mu^2}{x_1 Q^2 e^{\gamma_E}}+2 \right ]H^{(0)} \, ,
\label{eq:17}\\
G^{(1)}_{3c}&=&-\frac{\alpha_s}{8\pi N_c}
\left [\frac{1}{\varepsilon}+\ln\frac{4\pi \mu^2}{\delta_{12} Q^2 e^{\gamma_E}}-
\ln{\frac{\delta_{12}}{\delta_1}}\ln{\frac{\delta_{12}}{\delta2}}+\ln{\frac{\delta^2_{12}}{\delta_1 \delta_2}}+\frac{3}{2}-\frac{\pi^2}{3} \right ]H^{(0)} \, ,
\label{eq:18}\\
G^{(1)}_{3d}&=&\frac{\alpha_s N_c}{8\pi}
\left [\frac{3}{\varepsilon}+3\ln\frac{4\pi \mu^2}{\delta_{12} Q^2 e^{\gamma_E}}+
\ln{\frac{\delta_{12}}{\delta_2}}+2\ln{\frac{\delta_{12}}{\delta_1}}+\frac{11}{2}\right ] H^{(0)} \, ,
\label{eq:19}\\
G^{(1)}_{3e}&=&\frac{\alpha_s N_c}{8\pi}\left [\frac{3}{\varepsilon}+3\ln\frac{4\pi \mu^2}{x_{1} Q^2 e^{\gamma_E}}+
\ln(\frac{x_1}{\delta_2})(1-\ln\frac{x_1}{\delta_{12}})+\frac{1}{2} \ln{\frac{x_1}{\delta_{12}}}-\frac{2}{3} \pi^2+\frac{11}{2} \right ]
H^{(0)} \, .
\label{eq:20}
\eeq}\\
We give a discussion about $G^{(1)}_{3b}$ here.
Contrasting to the amplitudes of other diagrams which can be understood by the general IR analysis,
the calculation of Fig.~\ref{fig:fig3}(b)  does not generate IR divergence.
To explain this "anomaly", we should go back to the perturbative QCD factorization\cite{ChengRRA},
this IR piece is kinematic forbidden due to the initial and final spin structure we are interested.
Because the box and pentagon correction is UV safe, we sum up all the UV terms to see the coefficient
$\alpha_s/4\pi(11-2N_f/3)$,  which agrees with the universality of wave function in
Ref.~\cite{LiNN,ChengGBA}.

\begin{figure}[htbp]
\begin{center}
\vspace{0cm}
\includegraphics[width=0.6\textwidth]{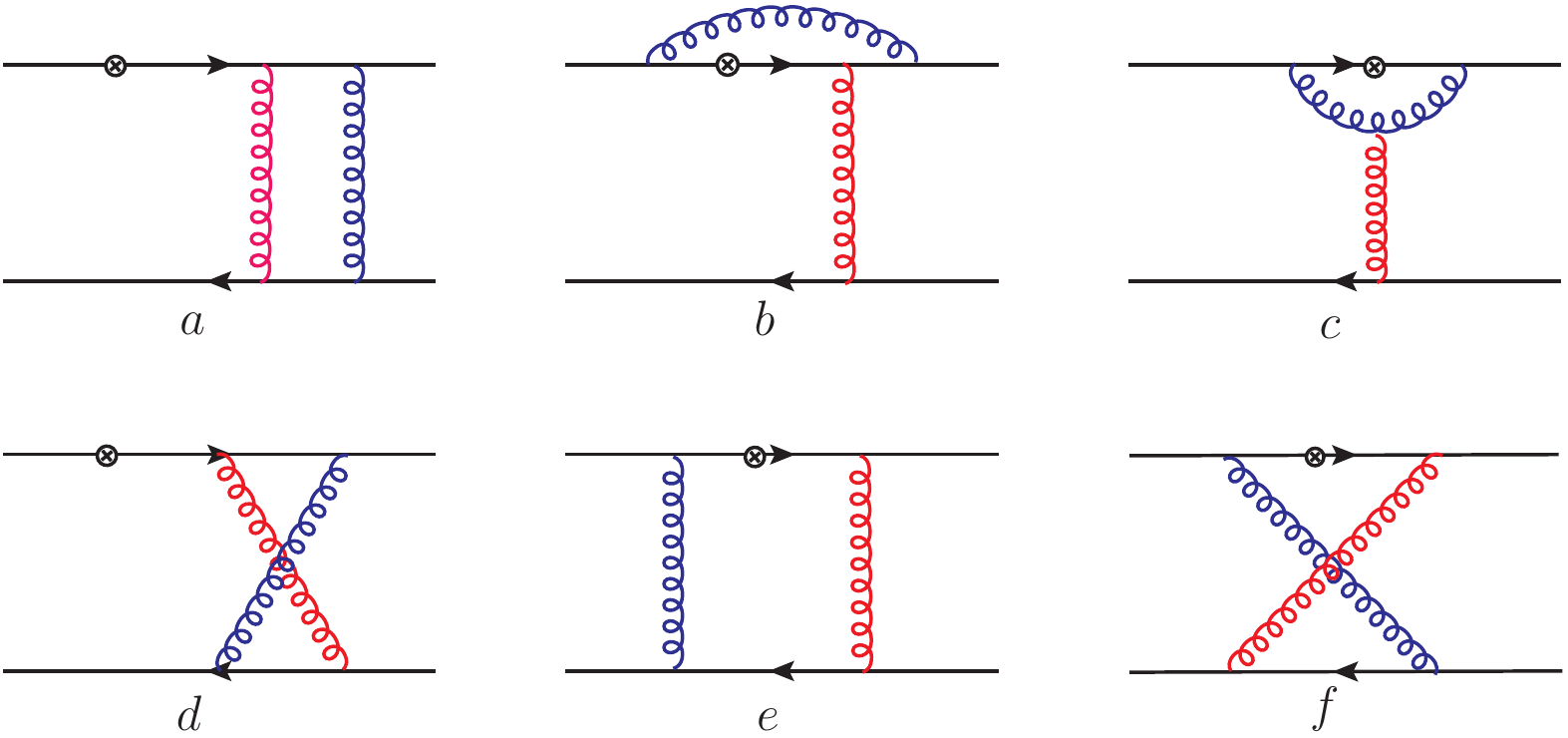}
\vspace{0.5cm}
\caption{Box and pentagon corrections to Fig.~\ref{fig:fig1}(b).}
\label{fig:fig4}
\end{center}
\end{figure}


The corrections from the box and pentagon diagrams in Fig.~\ref{fig:fig4} are arranged as:
{\small
\beq
G^{(1)}_{4b}&=&\frac{\alpha_s}{8\pi N_c}\left [\ln{\delta_1}\ln{\delta_2}-\ln{\delta_1}\ln{x_1}
-\ln{x_1}(1-\ln{x_1})+\ln\delta_2+\frac{\pi^2}{6}\right ] H^{(0)} \, , \label{eq:21}\\
G^{(1)}_{4c}&=& 0 \,, \label{eq:22}\\
G^{(1)}_{4d}&=&-\frac{\alpha_s}{8\pi N_c}\left [\ln{\frac{\delta_{12}}{\delta_1}}(\ln{\frac{x_1}{\delta_2}}+1)+\frac{\pi^2}{6} \right]
H^{(0)} \, , \label{eq:23}\\
G^{(1)}_{4f}&=&-\frac{\alpha_s}{8\pi N_c}\left [\ln{\frac{\delta_{12}}{x_1 \delta_1}}
\ln{\frac{\delta_{12}}{\delta_2}}+\frac{\pi^2}{4}-\frac{1}{2} \right ] H^{(0)} \, .
\label{eq:24}
\eeq}\\
We do not write down the results of reducible Fig.~\ref{fig:fig4} (a) and \ref{fig:fig4}(e)
since they cancel with their partner effective diagrams exactly.
Fig.~\ref{fig:fig4}(c) gives collinear logarithm $\ln\delta_1$ at the first sight,
but this IR piece is power suppressed by $ \Lambda^2_{QCD}/Q^2 $\cite{ChengGBA}.
We do not write down the adjoint correction to another LO kernel, obtained with replacing $x_1 \rightarrow x_2, \, k_{1T} \rightarrow k_{2T} $ from $H^{(0)}$,
in Eq.(~\ref{eq:24}) for Fig.~\ref{fig:fig4}(f).
We found that all double logs in Figs.~\ref{fig:fig3}(c) and ~\ref{fig:fig4}(b,d,f)
cancel each other due to the soft dynamics, rather different from the cases for the collinear light cone
wave functions.

\subsubsection{Effective diagrams at NLO}

In this section, we present the calculation of effective diagrams in terms of the convolution integration between NLO initial and final mesons wave functions and LO hard kernel.
To reproduce the collinear divergence in full amplitude, we focus on the hadronic matrix elements of wave function accordingly,
says leading transversal Fock states of $\rho$ meson and sub-leading valence Fock states with pseudoscalar current of $\pi$ meson.
{\small
\beq
\Phi_{\rho}^{T}(x_1,k_{1T};x'_1,k'_{1T})&=&\int \frac{dy^{-}}{2\pi}\frac{d^{2}y_T}{(2\pi)^{2}}
e^{-ix^{'}_1p^{+}_1y^{-}+i \mathbf{k^{'}_{1T}} \cdot \mathbf{y_T}} \non
&& \cdot \langle 0 \vert \bar{q}(y)\gamma_T \vsl W_y^{\dagger}(n_1) I_{n_1;y,0} W_0(n_1) q(0) \vert \bar{u}(p_1-k_1) d(k_1) \rangle \, ,
\label{eq:25}\\
\Phi_{\pi}^{T}(x'_2,k'_{2T};x_2,k_{2T}) &=& \int \frac{dz^{+}}{2\pi}\frac{d^{2}z_T}{(2\pi)^{2}}
e^{-ix^{'}_2p^{-}_2z^{+}+i \mathbf{k^{'}_{2T}} \cdot \mathbf{z_T}} \non
&& \cdot \langle 0 \vert \bar{q}(y)\gamma_5 W_z^{\dagger}(n_2) I_{n_2;z,0} W_0(n_2) q(0) \vert \bar{u}(p_2-k_2) d(k_2) \rangle \, ,
\label{eq:26}
\eeq}\\
where $y=(0,y^-,\mathbf{y_T})$ and $z=(z^+,0,\mathbf{z_T})$ are light cone coordinates of the anti-quark field.
Wilson lines are defined with a litter bit straying from the light cone, $n_1^2,n_2^2 \ne 0$:
{\small
\beq
W_y(n_1) &=& \mathrm{P \, \exp}\left[{-ig_s \int_0^\infty d\lambda n \cdot A(y+\lambda n_1)}\right] \, ,
\label{eq:27}\\
W_z(n_2) &=& \mathrm{P \, \exp}\left[{-ig_s \int_0^\infty d\lambda v \cdot A(z+\lambda n_2)}\right] \, ,
\label{eq:28}
\eeq}\\
in which P is the path-ordering operator and their nonzero order terms redistribute the momentum between the meson institutes.
Wilson lines at two different points are connected by a vertical link at infinity \cite{JiAA}.
In this way, we can evade the light cone singularity ($l \parallel n/v$) by the scalar regulators $\xi_1^2 \equiv 4(n_1 \cdot p_1)^2/\vert n_1^2 \vert$
and $\xi_2^2 \equiv 4(n_2 \cdot p_2)^2/\vert n_2^2 \vert$.
This rapidity singularity had been investigated by the joint resummation \cite{LiXNA}
and the result shows the scheme-dependence is small,
so in this paper we fix $\xi_1^2 = \xi_2^2 = Q^2$ to minimize the scheme dependence.

\begin{figure}[htbp]
\begin{center}
\vspace{0cm}
\includegraphics[width=0.6\textwidth]{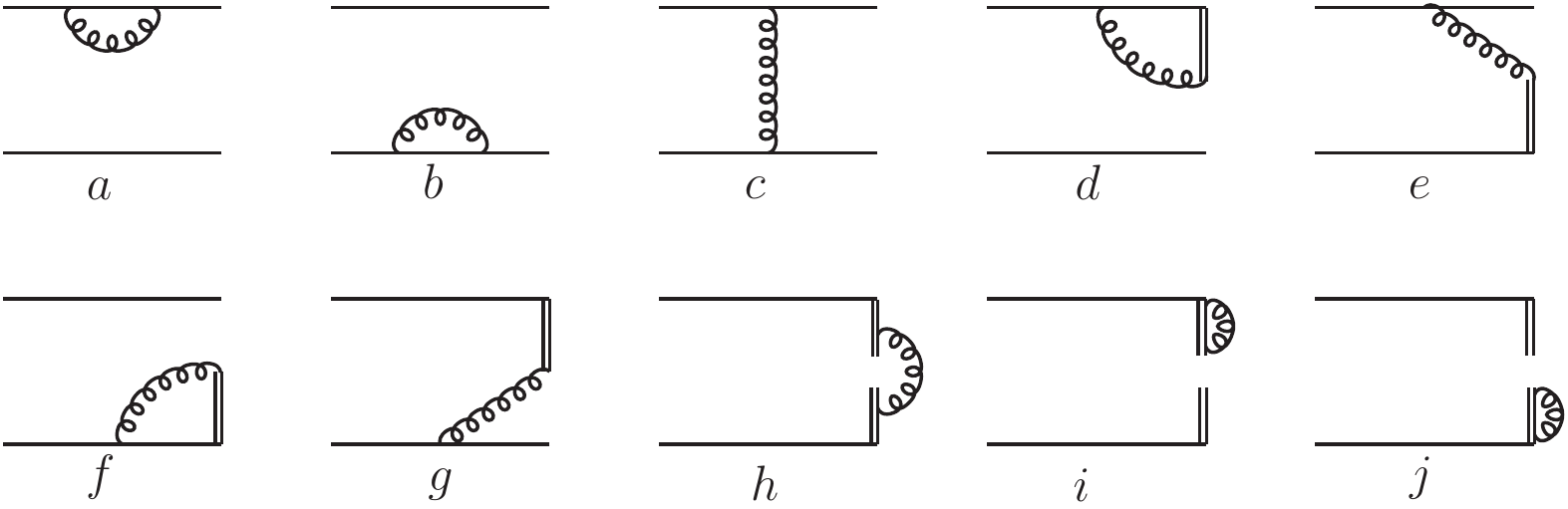}
\vspace{0.5cm}
\caption{The effective diagrams for the initial $\rho$ meson wave function.}
\label{fig:fig5}
\end{center}
\end{figure}


We firstly consider the second term in the right hand side (RHS) of Eq.~(\ref{eq:10}),
where the NLO wave function of initial state meson can be obtained from Eq.~(\ref{eq:25}) with the 1st-order expansion of Wilson line in Eq.~(\ref{eq:27}).
Effective Feynman diagrams of NLO wave function $\Phi_\rho^T$ with eikonal propagator indicating in double line is depicted in Fig.~\ref{fig:fig5},
and we calculate the convoluted integral
{\small
\beq
\Phi^{(1)}_\rho \otimes \mathcal{H}^{(0)} \equiv \int dx'_1 d^{2} \mathbf{k}'_{1T} \, \Phi^{T,\,(1)}_\rho(x_1,\mathbf{k}_{1T};x'_1,\mathbf{k}'_{1T})
\, \mathcal{H}^{(0)}(x'_1,\mathbf{k}'_{1T};x_2,\mathbf{k}_{2T}),
\label{eq:29}
\eeq}\\
and one by one,
{\small
\beq
\Phi^{(1)}_{\rho, a} {\otimes} \mathcal{H}^{(0)} &=& \Phi^{(1)}_{\rho,b} \small{\otimes} H^{(0)} =
-\frac{\alpha_s C_F}{8\pi}\left (\frac{1}{\varepsilon}+\ln\frac{4\pi \mu^2_f}{\delta_1 Q^2 e^{\gamma_E}}+2 \right) \, H^{(0)} \,,
\label{eq:30}\\
\Phi^{(1)}_{\rho,c} {\otimes} \mathcal{H}^{(0)} &=& 0 \,,
\label{eq:31}\\
\Phi^{(1)}_{\rho,d} {\otimes} \mathcal{H}^{(0)} &=&\frac{\alpha_s C_F}{4\pi}
\left (\frac{1}{\varepsilon}+\ln\frac{4\pi \mu^2_f}{k^2_{1T} e^{\gamma_E}}
-\ln^2 \frac{\xi^2_1}{k^2_{1T}}+\ln\frac{\xi^2_1}{k^2_{1T}} +2-\frac{\pi^2}{3} \right ) \, H^{(0)} \, ,
\label{eq:32}\\
\Phi^{(1)}_{\rho,e} {\otimes} \mathcal{H}^{(0)} &=&\frac{\alpha_s C_F}{4\pi}
\left (\ln^2 \frac{x_1 \xi^2_1}{k^2_{1T}}+\frac{2 \pi^2}{3} \right ) \, H^{(0)}  \, ,
\label{eq:33}\\
\Phi^{(1)}_{\rho,f} {\otimes} \mathcal{H}^{(0)} &=&\frac{\alpha_s C_F}{4\pi}\left (\frac{1}{\varepsilon}
+\ln\frac{4\pi \mu^2_f}{k^2_{1T} e^{\gamma_E}}
-\ln^2 \frac{x^2_1 \xi^2_1}{k^2_{1T}}+\ln\frac{x^2_1 \xi^2_1}{k^2_{1T}} +2-\frac{\pi^2}{3} \right ) \, H^{(0)} \, ,
\label{eq:34}\\
\Phi^{(1)}_{\rho,g} {\otimes} \mathcal{H}^{(0)} &=&\frac{\alpha_s C_F}{4\pi}
\left (\ln^2 \frac{x^2_1 \xi^2_1}{k^2_{1T}}-\frac{\pi^2}{3} \right ) \, H^{(0)}  \,,
\label{eq:35}\\
\Phi^{(1)}_{\rho,h} {\otimes} \mathcal{H}^{(0)} &=&\frac{\alpha_s C_F}{2\pi}
\left (\frac{1}{\varepsilon}+\ln\frac{4\pi \mu^2_f}{\delta_{12} Q^2 e^{\gamma_E}} \right) \, H^{(0)} \, ,
\label{eq:36}
\eeq}\\
with the factorization scale $\mu_f$.
We can also see that the double log $\ln^2{k_T}$ disappears ultimately due to the same reason as in the full amplitudes.
We naively consider the reducible Fig.~\ref{fig:fig5}(c) as zero because it also reproduce the result of quark diagram Fig.~\ref{fig:fig4}(e) exactly.
Their summation gives:
{\small
\beq
\sum_{i=a,\cdots,h} \Phi^{(1)}_{\rho,i} \otimes \mathcal{H}^{(0)}
&=& \frac{\alpha_s C_F}{4\pi} \left[\frac{3}{\varepsilon}
+3\ln\frac{4\pi \mu^2_f}{\xi_1Q^2e^{\gamma_E}}+(2\ln x_1+4)\ln\frac{\xi^2_1}{\delta_1 Q^2} \right. \non
&~& \left. \hspace{1cm} +2\ln\frac{\xi^2_1}{\delta_{12}Q^2}+\ln x_1(\ln x_1+2)
+2-\frac{\pi ^2}{3}\right] \, H^{(0)}.
\label{eq:37}
\eeq}

We also calculate the third term in the RHS of Eq.~(\ref{eq:10}), with the wave function of
final state meson is Eq.~(\ref{eq:26}).
{\small
\beq
\mathcal{H}^{(0)} \otimes \Phi^{(1)}_\pi = \int dx'_2 d^{2}\mathbf{k}'_{2T}  \,
H^{(0)}(x'_1,\mathbf{k}'_{1T};x_2,\mathbf{k}_{2T}) \, \Phi^{P, \,(1)}_\pi(x_2,\mathbf{k}_{2T};x'_2,\mathbf{k}'_{2T})
\label{eq:38}
\eeq}\\
The effective Feynman diagrams for the NLO wave function of final state are  similar with those
as shown in Fig.~\ref{fig:fig5},
we do not show the details of them for the conciseness, and only show the summed result
{\small
\beq
\sum_{i=a,\cdots,h} \mathcal{H}^{(0)} \otimes \Phi^{(1)}_{\pi,i} &=& \frac{\alpha_s C_F}{8\pi} \left[\frac{2}{\varepsilon}+2\ln\frac{4\pi \mu^2_f}{\xi_2Q^2e^{\gamma_E}}
+(2\ln x_2+4)\ln\frac{\xi^2_2}{\delta_2 Q^2} \right. \non
&~& \left. \hspace{1cm} + 2\ln\frac{\xi^2_2}{\delta_{12}Q^2}+\ln x_2(\ln x_2+2)-\frac{\pi ^2}{3}\right]  \, H^{(0)} \, .
\label{eq:46}
\eeq}\\
The results of irreducible effective amplitudes in Eq.~(\ref{eq:46}) is half of that in Eq.~(\ref{eq:36}) due to the different spin structures in wave functions,
which lead to the different UV behaviour and the half collinear divergence.

\subsubsection{NLO hard correction}

Before extracting the NLO form factors, we firstly confirm the IR cancelation between
the quark diagrams and the effective diagrams.
Taking into account of the jet function effect, which emerged when the internal quark is on-shell in the small $x_1$ region\cite{LiAY},
{\small
\beq
J^{(1)} \, H^{(0)}=-\frac{1}{2} \frac{\alpha_s(\mu_f)C_F}{4\pi} \left[\ln^2x_1+\ln x_1+\frac{\pi^2}{3}\right] \, H^{(0)} \, ,
\label{eq:47}
\eeq}\\
we obtain the NLO hard kernel 
in the $\overline{\mathrm{MS}}$ scheme with Eq.~(\ref{eq:10}) ,
{\small
\beq
&~&H^{(1)}(\mu, \mu_f, Q^2)  \rightarrow H^{(1)} - J^{(1)} \, H^{(0)} \equiv \mathcal{F}_{\rho\pi}^{(1)}(\mu, \mu_f, Q^2) \, H^{(0)} \non
&=& \frac{\alpha_s(\mu_f) C_F}{8\pi} \left[
\frac{21}{2}\ln{\frac{\mu^2}{Q^2} -8\ln{\frac{\mu_f^2}{Q^2}}} + \frac{9}{4}\ln{x_1}\ln{x_2}  - \frac{3}{4} \ln^2{x_1} - \ln^2{x_2} \right. \non
&& \left. \hspace{1.7cm} - \frac{67}{8}\ln{x_1} - 2 \ln{x_2} + \frac{37}{8} \ln{\delta_{12}} + \frac{107}{8} - \frac{\pi^2}{3} \right] \, H^{(0)} \,\,\, ,
\label{eq:48}
\eeq}\\
where the $k_T$ independent function $\mathcal{F}_{\rho\pi}^{(1)}(Q^2)$ is the NLO correction to spacelike rho-pion from factor.

\subsection{Derivation of the timelike rho-pion form factor at NLO}

To obtain the NLO timelike rho-pion form factor,
we recall the exchanging symmetry $-x_1 \leftrightarrow x_1$ between spacelike and timelike
form factor in PQCD approach as we shown at LO.
We do not do the complicate NLO calculations again,
what we suggest is to take the NLO result of spacelike form factor obtained in the above
subsection, and then make a analytic continuation to the timelike region \cite{HuCP,ChengQRA}.
We use the following continuation prescriptions
\beq
 \ln{Q^2} &\rightarrow & \ln{(-Q^2 - i \epsilon)} = \ln{Q^2} - i \pi \, ,
\label{eq:50} \\
  \ln{x_1} &=& \ln{\frac{-x_1Q^2 + k_{1T}^2 + i\epsilon}{Q^2+i \epsilon}} = \ln{\frac{-x_1Q^2 + k_{1T}^2+ i\epsilon}{Q^2}} - i	\pi
\equiv \ln{x'_1} - i\pi \, ,
\label{eq:51}\\
\ln{\delta_{12}} &=& \ln{\frac{-x_1x_2Q^2 + |k_{1T}+k_{2T}|^2 + i\epsilon}{Q^2+i \epsilon}} 
= \ln{\frac{-x_1x_2Q^2 + |k_{1T}+k_{2T}|^2+ i\epsilon}{Q^2}}   - i\pi  \non
&\equiv& \ln{\delta'_{12}} - i\pi \, ,
\label{eq:52}
\eeq
and take fourier transfermation of the transversal momentum appeared above to its conjugate
coordinate space. In order to be consistent in form with those formula at LO,
the NLO correlation function to timelike rho-pion form factor can be written as,
{\small
\beq
&&\mathcal{G}^{(1)}_{\rho\pi}(\mu,\mu_f,Q^2,b_{i}) = \frac{\alpha_s(\mu_f) C_F}{8\pi} \left\{ \left[
\frac{21}{2}\ln{\frac{\mu^2}{Q^2}  -8\ln{\frac{\mu_f^2}{Q^2}}} + \frac{9}{4}\left(\frac{1}{2}\ln{\frac{4x_1}{Q^2b_1^2}}-\gamma_E  \right) \ln{x_2}  \right. \right. \non
&& \left.\left. \hspace{3.5cm}  - \frac{3}{4} \left(\frac{1}{2}\ln{\frac{4x_1}{Q^2b_1^2}}-\gamma_E \right)^2
- \frac{67}{8} \left(\frac{1}{2}\ln{\frac{4x_1}{Q^2b_1^2}}-\gamma_E\right)  \right.\right. \non
&& \left.\left. \hspace{3.5cm}  + \frac{37}{8} \left(\frac{1}{2}\ln{\frac{4x_1x_2}{Q^2b_1^2}}-\gamma_E\right)
-\ln^2{x_2} -2 \ln{x_2}  + \frac{65\pi^2}{48} + \frac{107}{8} \right]  \right. \non
&& \left. \hspace{3.5cm} + i\pi \left[\frac{9}{4}\left(\frac{1}{2}\ln{\frac{4x_1}{Q^2b_1^2}}-\gamma_E\right)
- \frac{27}{8} \ln{x_2} + \frac{25}{8} \right] \right\} \, .
\label{eq:53}
\eeq}\\

\section{NUMERICAL ANALYSIS}

\subsection{PQCD prediction}

We preform the numerical analysis in this section, the rho-pion transition form factor up to
NLO is derived as,
{\small
\beq
&&F_{\rho\pi}(Q^2)=\frac{64\pi}{9}  \alpha_s(\mu_f) \int_0^1 dx_1dx_2 \int_0^\infty b_1db_1 b_2db_2 \, \textrm{exp}[-S_{\rho\pi}(x_i,b_i,Q,\mu)] \non
&&\hspace{2cm} \left\{ m_\rho (\phi_\rho^v(x_1)-\phi_\rho^a(x_1)) \phi_\pi^A(x_2) h(x_2,x_1,b_2,b_1) \right. \non
&&\left. \hspace{2cm} + x_1 m_\rho (\phi_\rho^v(x_1)-\phi_\rho^a(x_1)) \phi_\pi^A(x_2) h(x_1,x_2,b_1,b_2) \right. \non
&&\left. \hspace{2cm} + 2m_0^\pi \phi_\rho^T(x_1) \phi^P_\pi(x_2) \left[1 + \mathcal{F}^{(1)}_{\rho\pi}(\mu,\mu_f,Q^2)\right] h(x_1,x_2,b_1,b_2)  \right\}  S_t(x_1) S_t(x_2)\,,\,\,\,
\label{eq:54}
\eeq}
{\small
\beq
&&G_{\rho\pi}(Q^2)=\frac{64\pi}{9}  \alpha_s(\mu_f) \int_0^1 dx_1dx_2 \int_0^\infty b_1db_1 b_2db_2 \, \textrm{exp}[-S_{\rho\pi}(x_i,b_i,Q,\mu)] \non
&&\hspace{2cm} \left\{ m_\rho (\phi_\rho^v(x_1)-\phi_\rho^a(x_1)) \phi_\pi^A(x_2) h'(x_2,x_1,b_2,b_1) \right. \non
&&\left. \hspace{2cm} - x_1 m_\rho (\phi_\rho^v(x_1)-\phi_\rho^a(x_1)) \phi_\pi^A(x_2) h'(x_1,x_2,b_1,b_2) \right. \non
&&\left. \hspace{2cm} + 2m_0^\pi \phi_\rho^T(x_1) \phi^P_\pi(x_2) \left[1+ \mathcal{G}^{(1)}_{\rho\pi}(\mu,\mu_f,Q^2,b_{i}) \right] h'(x_1,x_2,b_1,b_2)  \right\}  S_t(x_1) S_t(x_2)\,.\,\,\,
\label{eq:55}
\eeq}\\
in which the light cone distribution amplitudes (LCDAs) are taken up to $n=2$
and $n=4$ in the Gegenbauer expansion of rho and pion meson, respectively,
{\small
\beq
\phi^T_\rho(x)&=&\frac{f_\rho^T}{\sqrt{2N_C}} x (1-x) \left[ 1 + a_{2,\rho}^\perp C_2^{3/2}(t) \right] \, ,
\label{eq:56}\\
\phi^v_\rho(x)&=&\frac{f_\rho}{2\sqrt{2N_C}} \left[ \frac{3}{4}(1+t^2) + \left(\frac{3}{7} a_{2,\rho}^\parallel + 5 \zeta_3^A\right) (3t^2-1) \right] \, ,
\label{eq:57}\\
\phi^a_\rho(x)&=&\frac{3f_\rho}{2\sqrt{2N_C}}(1-2x) \left\{ 1 + 4\left[ \frac{1}{4}a_{2,\rho}^\parallel+ \frac{5}{3} \zeta_3^A \left( 1-\frac{3}{16}\omega_{1,0}^A\right)
+ \frac{35}{4} \zeta_3^V\right] (10x^2-10x+1) \right\} \, ,
\label{eq:58}\\
\phi^A_\pi(x)&=&\frac{3f_\pi}{\sqrt{2N_C}} x (1-x) \left[1+a_2^\pi C_2^{3/2}(t) + a_4^\pi C_4^{1/2}(t) \right] \, ,
\label{eq:59}\\
\phi^P_\pi(x)&=&\frac{f_\pi}{2\sqrt{2N_C}} \left[ 1 + \left(30 \eta_3 - \frac{5}{2} \rho_\pi^2 \right) C_2^{1/2}(t)
- 3 \left(\eta_3 \omega_3 + \frac{9}{20} \rho_\pi^2 (1 + 6 a_2^\pi)\right) C_4^{1/2}(t) \right] \, .
\label{eq:60}
\eeq} \\
To do the numerics, we firstly use the asymptotic DAs with only the lowest terms,
we also suggest to use another set of DAs to check the effects of high order Gegenbauer moments,
\begin{enumerate}
\item[] \,\,\,
$f_\rho^T = 0.160 \, \mathrm{GeV}, \,\,\, f_\rho = 0.216 \,\mathrm{GeV}, \, \, \, a_{2,\rho}^\perp = 0.14, \,\,\, a_{2,\rho}^\parallel = 0.17, $ \,\,\, \cite{StraubICA}
\vspace{-1em}
\item[] \,\,\,
$\zeta_3^A = 0.032, \,\,\, \zeta_3^V = 0.013, \,\,\, \omega_{0,1}^A = -2.1, $ \,\,\, \cite{BallSK}
\vspace{-1em}
\item[] \,\,\,
$a_2^\pi = 0.35, \,\,\, a_4^\pi = 0.12, $ \,\,\, \cite{ChengAA}
\vspace{-1em}
\item[] \,\,\,
$f_\pi = 0.130 \, \mathrm{GeV}, \,\,\, \rho_\pi = m_\pi/m_0^\pi = 0.139/1.4,\,\,\,
\eta_3 = 0.015, \,\,\, \omega_3 = -3.0 \, .$ \,\,\, \cite{DuplancicIX}
\end{enumerate}

In this work we concentrate on the evaluations for the central values of the relevant form factors
only and do not consider the effects of the
uncertainties, coming from the errors of the above parameters at certain scale
and those from their scale evolution, and from the choice of factorization and renormalization scale,
as well as some other scheme dependence.

\begin{figure}[htbp]
\begin{center}
\vspace{0.4cm}
\includegraphics[width=0.4\textwidth]{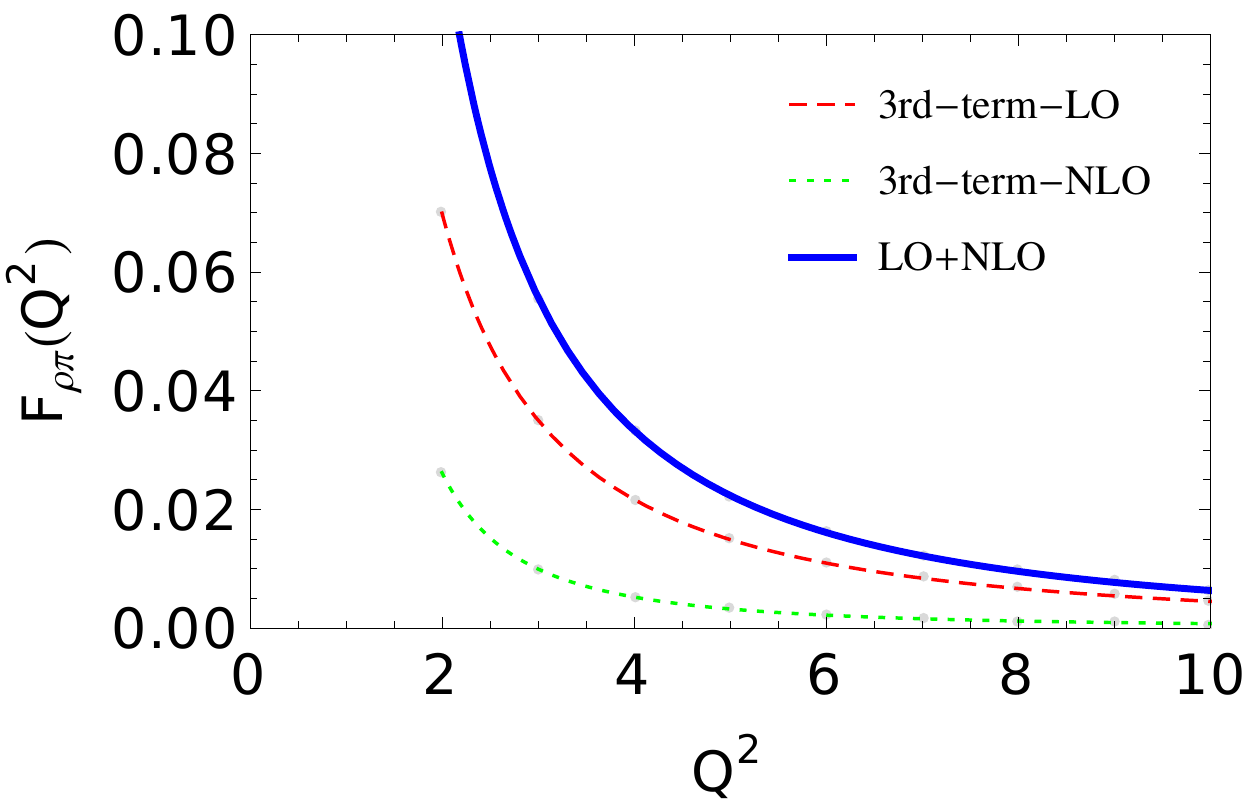}
\hspace{4mm}
\includegraphics[width=0.4\textwidth]{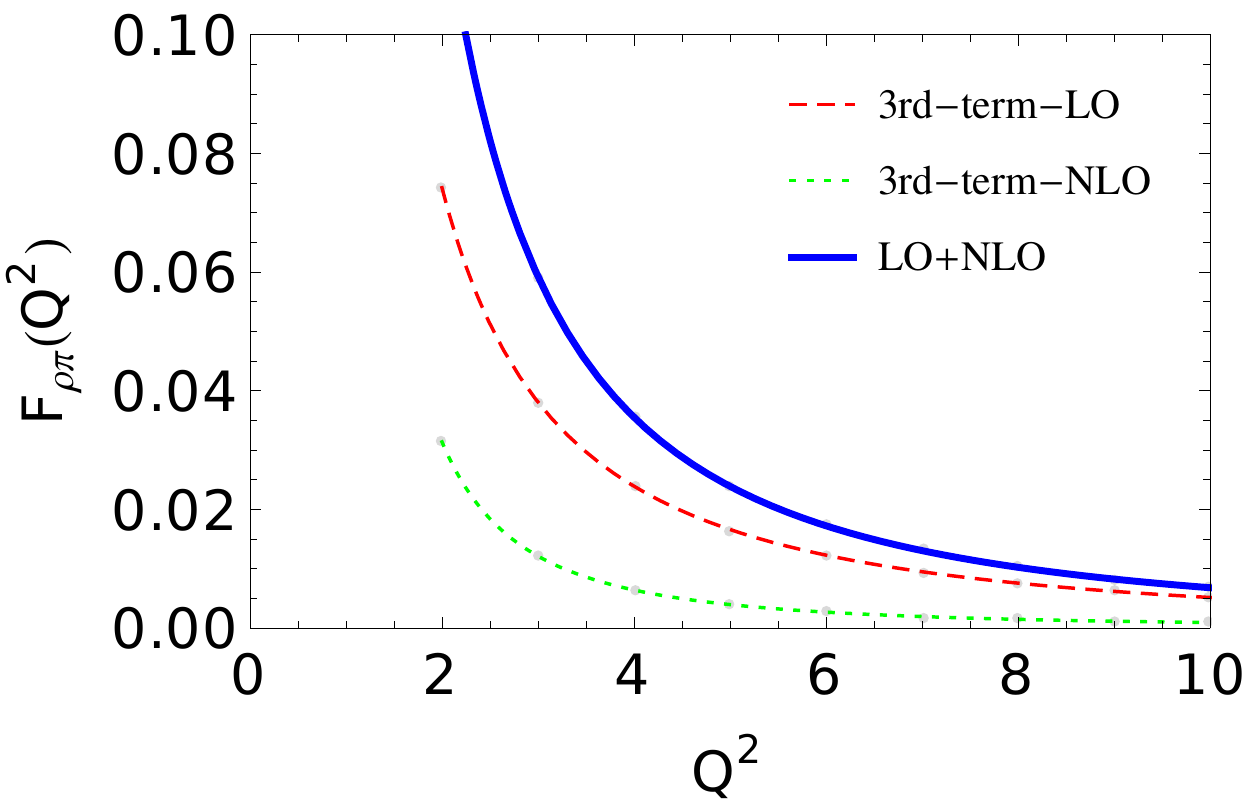}\\
\vspace{4mm}
\includegraphics[width=0.4\textwidth]{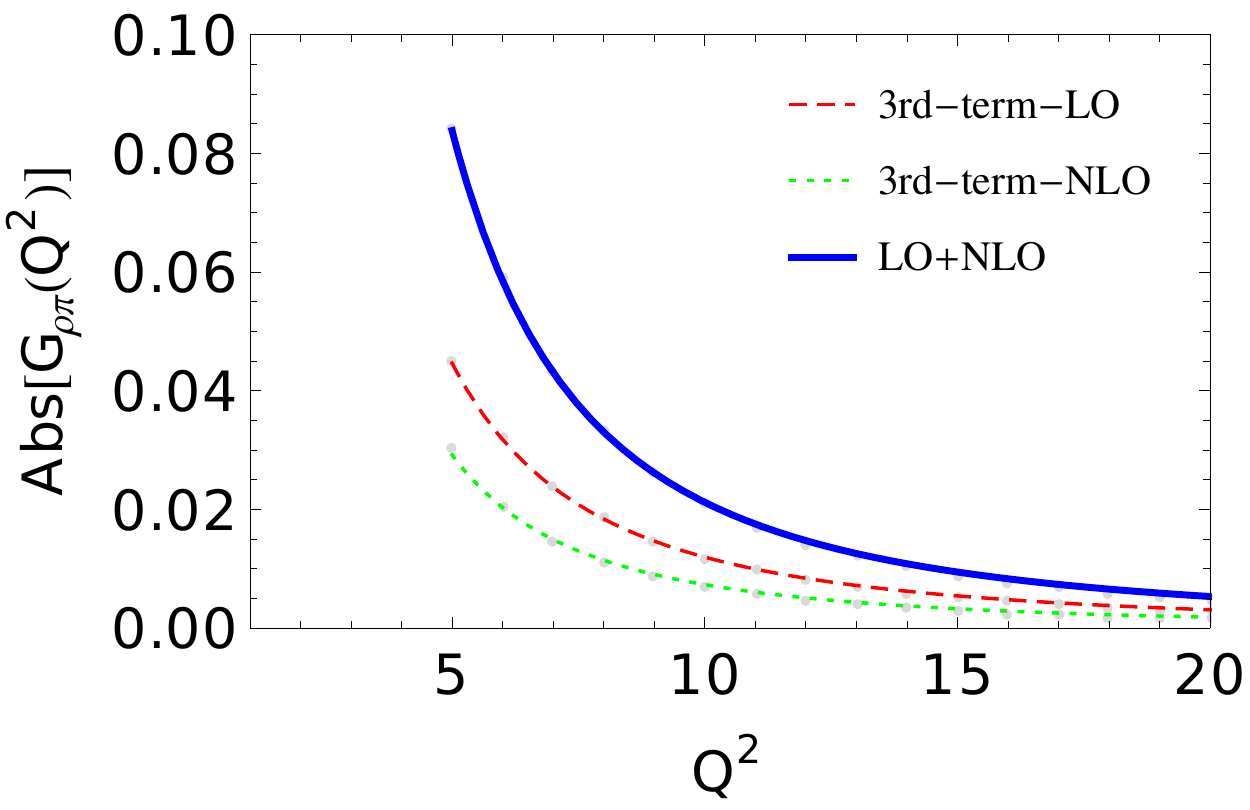}
\hspace{4mm}
\includegraphics[width=0.4\textwidth]{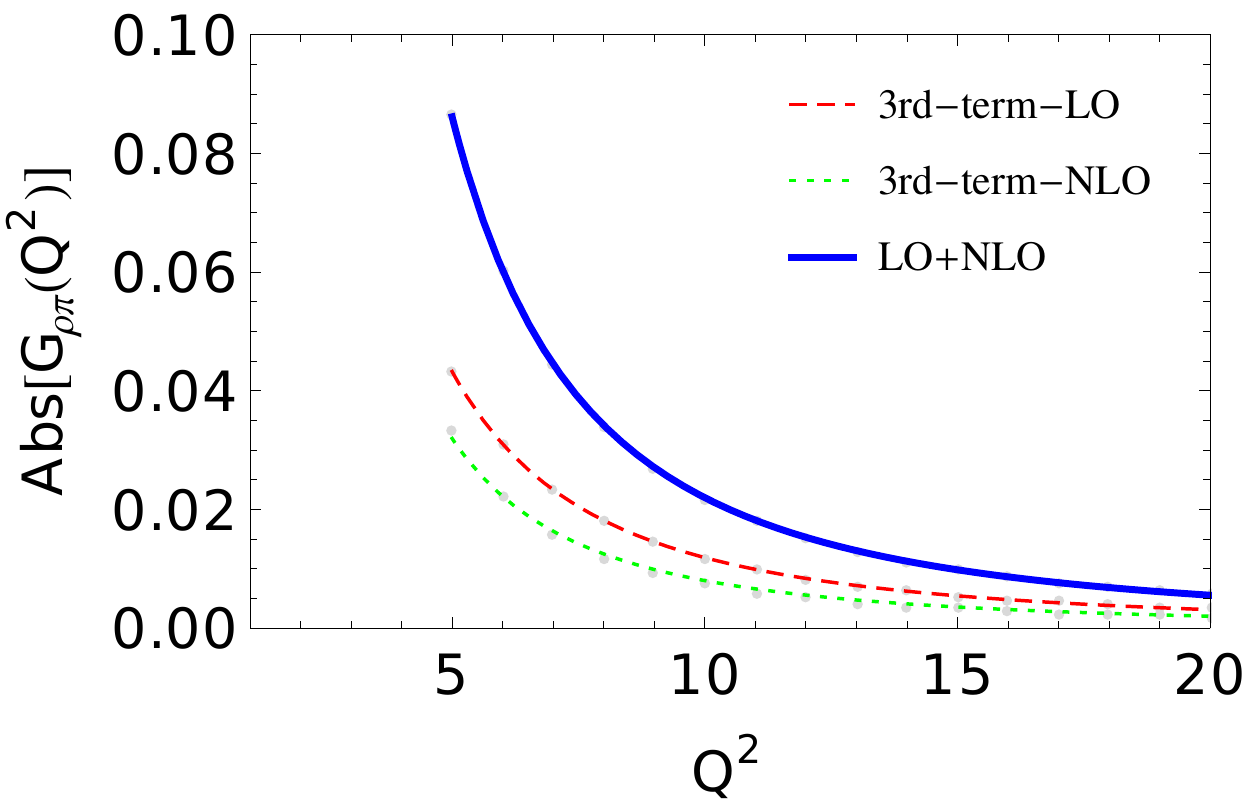}\\
\vspace{4mm}
\includegraphics[width=0.4\textwidth]{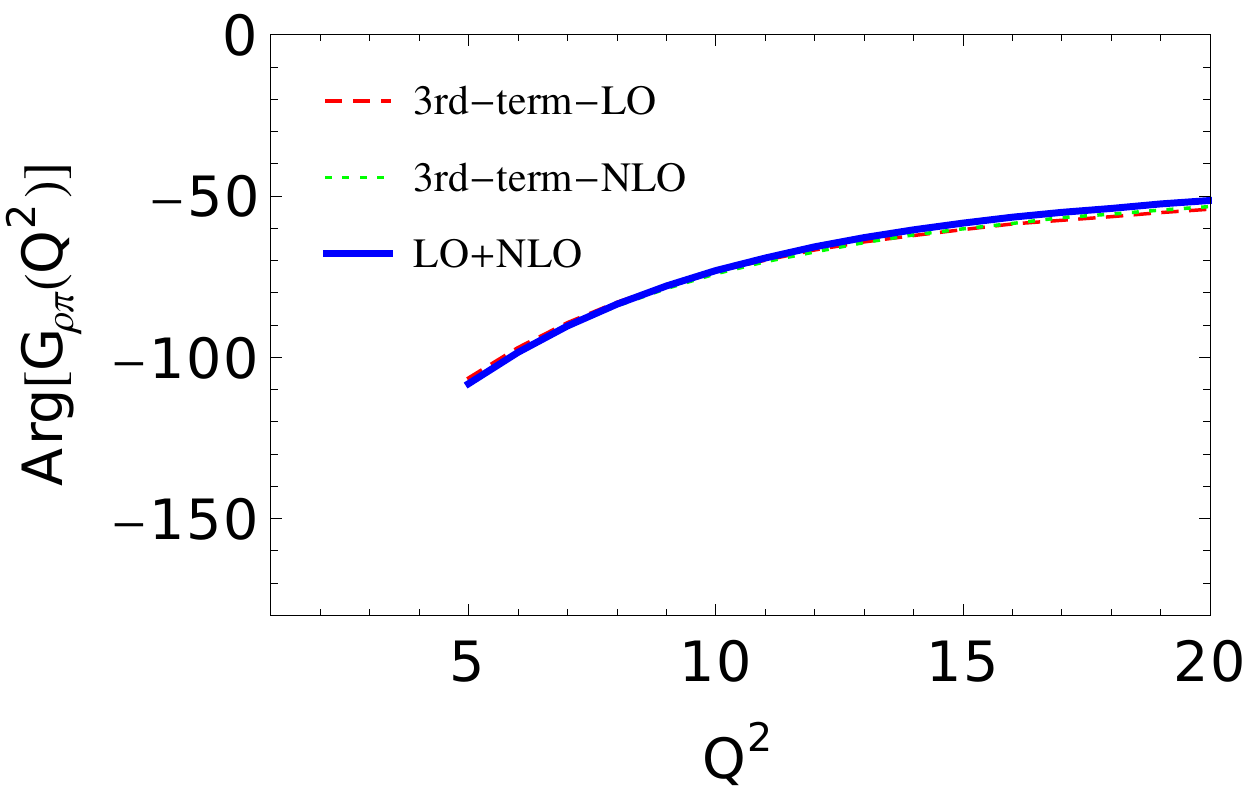}
\hspace{4mm}
\includegraphics[width=0.4\textwidth]{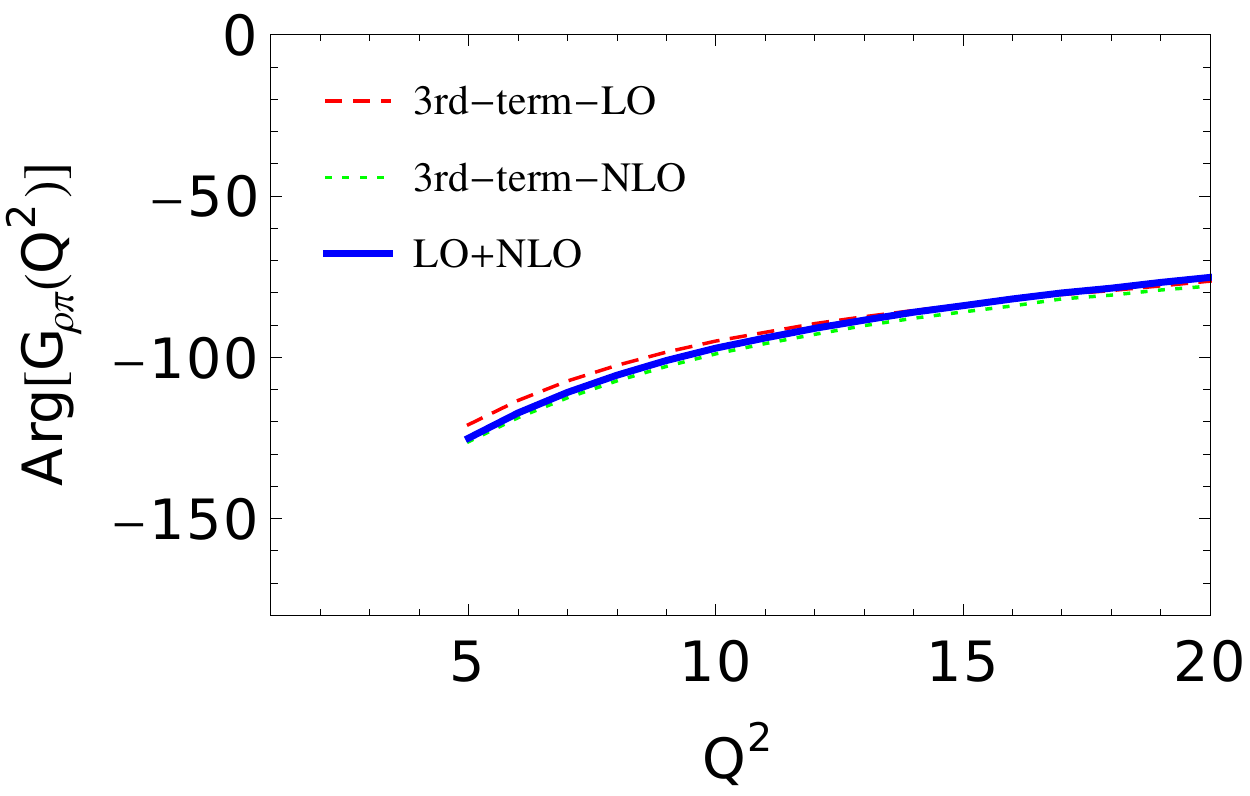}
\vspace{0.1cm}
\caption{Spacelike and timelike rho-pion transition form factors evaluated from PQCD with
the asymptotic (left) and nonasymptotic (right) rho and pion DAs.}
\label{fig:fig6}
\end{center}
\end{figure}

The PQCD predictions for the $Q^2$-dependeces of the form factors $F_{\rho\pi}(Q^2)$ and
$G_{\rho\pi}(Q^2)$ are depicted in Fig.~\ref{fig:fig6}, where the dominate contribution
term at LO and NLO, as well as the total results are exhibited.
When using the asymptotic DAs as input, we find that the NLO correction is less than $35\%$
in the spacelike region $Q^2 \ge 2 \, \mathrm{GeV}^2$ (the region PQCD applicable).
The timelike form factor is studied with the starting point $Q^2=5 \, \mathrm{GeV}^2$
since in the intermediate rho-pion invariant mass region PQCD fails to describe the resonant
mesons ( $\omega(782), \, \omega(1420)$ and $\omega(1650)$ ) \footnote{ The $\phi(1020), \,
\phi(1680)$ channel is mainly occupied by $K\bar{K^\ast}$.},
the convergence of the NLO correction to absolute value is not good before
$Q^2 \ge 10 \, \mathrm{GeV}^2$ (if we assume the convergence means $\le 50 \%$),
while the NLO correction retains the shape of strong phase.
Including the high Gegenbauer expansion terms brings a litter bit change to the results
when only the asymptotic DAs are taken into account,
which provides us an independent opportunity to determine the moments if
precision data become available.

\subsection{Interplaying with the lattice result}

\begin{figure}[htbp]
\begin{center}
\vspace{0.4cm}
\includegraphics[width=0.4\textwidth]{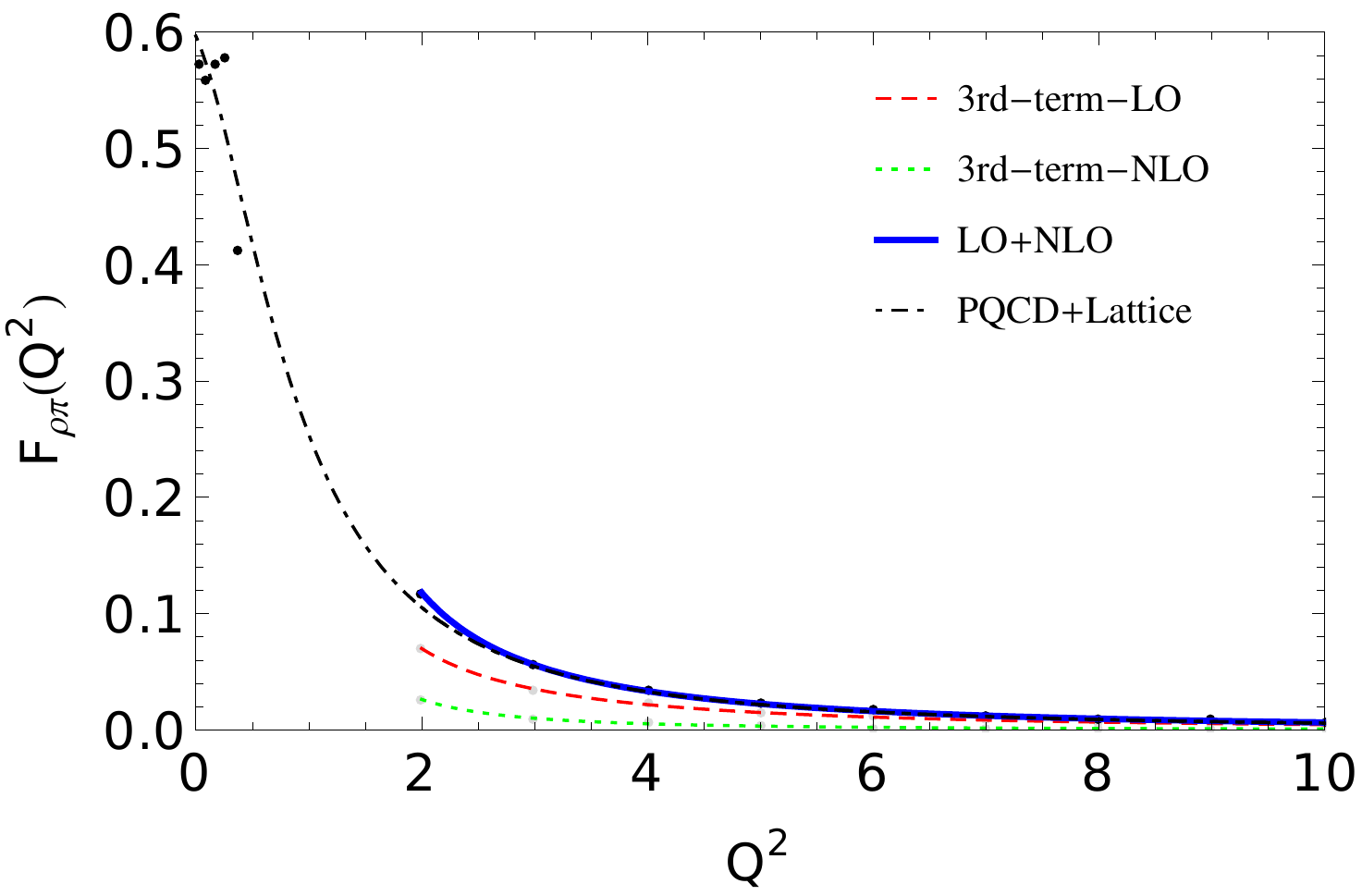}
\hspace{4mm}
\includegraphics[width=0.4\textwidth]{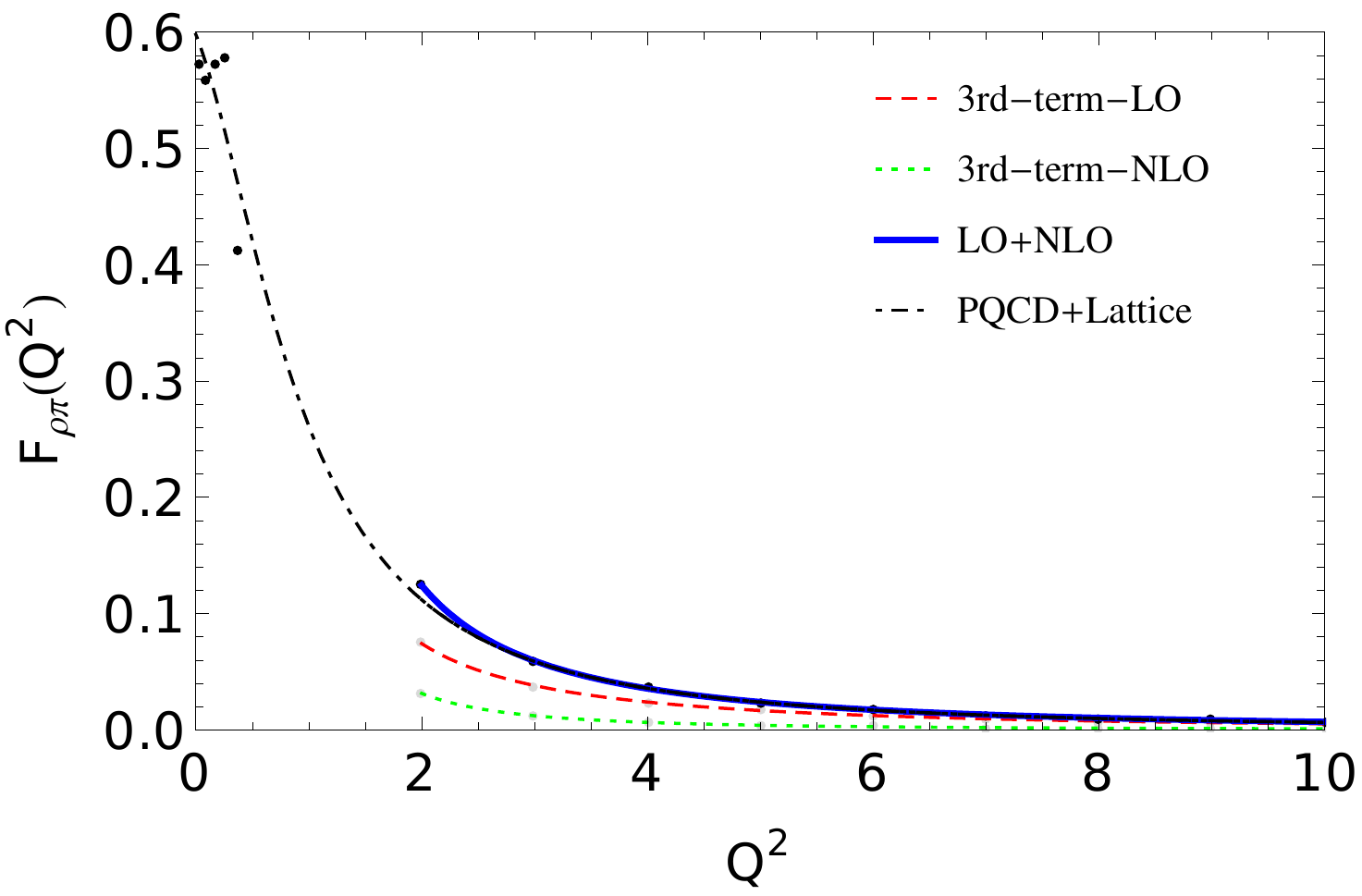}
\vspace{0.1cm}
\caption{Combine fitting of spacelike rho-pion form factor in PQCD and the Lattice QCD.
Left (right) plot shows the PQCD result obtained with asymptotic (nonasymptotic) DAs.}
\label{fig:fig7}
\end{center}
\end{figure}

Benefiting from the lattice QCD predictions at low $Q^2$ region \cite{OwenFRA},
we are able to known the rho-pion form factors at the the whole region of $Q^2$.
Because of the same reason of the broad resonance contribution in the intermediate timelike energy,
we are now only able to do the global fit for spacelike rho-pion form factor.
In Fig.~\ref{fig:fig7}, we show the spacelike result in the whole $Q^2$ region obtained by
combining fit of the PQCD predictions and the Lattice QCD evaluations.
Reciprocal of square polynomial parameterization is adopt \cite{KhodjamirianTK},
{\small
\beq
F_{\rho\pi}(Q^2) = \frac{A_{\rho\pi}}{Q^4 + Q^2 B_{\rho\pi} + C_{\rho\pi}},
\eeq}\\
with the use of the asymptotic (nonasymptotic) DAs,
we find numerically $A_{\rho\pi} = 0.606(0.676), \, B_{\rho\pi} = 0.370(0.457), \, C_{\rho\pi} = 1.016(1.131)$,
and $g_{\rho\pi\gamma^\ast} = F_{\rho\pi}(0) = A_{\rho\pi}/C_{\rho\pi} = 0.596(0.598)$,
which, within the range of the possible theoretical uncertainties, are consistent with currently
available data \cite{HustonWI,JensenNF,CapraroRP}.
As an by product, we can also estimate the charged rho-pion radius in this way
$\langle r_{\rho\pi}^2 \rangle = 1.304(1.449) \, \mathrm{GeV}^{-2}$.

\section{CONCLUSION}

In this paper, the rho-pion transition from factors  $F_{\rho\pi}(Q^2)$ and
$G_{\rho\pi}(Q^2)$  are studied with the inclusion of the QCD corrections at NLO in the framework
of the PQCD factorization approach.

We firstly calculate all the quark diagrams of the spacelike form factor as well as their
descendent effective diagrams that absorb all the residual collinear singularities,
and then take their difference to obtain the NLO hard corrections.
The spacelike form factor $F_{\rho\pi}(Q^2)$  is then extended  analytically to
the timelike one $G_{\rho\pi}(Q^2)$  based on the kinematic exchanging symmetry.
When adopting the asymptotic DAs of rho, pion mesons,
the NLO contribution provide a enhancement to  the LO result by less than $35\%$
for spacelike rho-pion form factor in the region $Q^2 \ge  2 \, \mathrm{GeV}^2$,
and the corresponding correction to the timelike form factor also support the perturbative
theory at large $Q^2$ region.

The recent Lattice QCD results in the low $Q^2$ region are  also used, together with the PQCD
predictions, to do the global fit for spacelike form factor in the whole energy extent,
and we get the rho-pion coupling $g_{\rho\pi\gamma^\ast} = 0.596$.
The combine fit is impossible now for timelike form factor due to the unclear
intermediate resonances in the broad medium energy region.


\section{ACKNOWLEDGEMENTS}

This work is supported by the National Natural Science Foundation of China under the 
No. 11235005 and 11775117.
S. Cheng gratefully acknowledge the NNU support during his visit when the analytical 
part of this project was finished.


\end{document}